# Volumetric glass modification with Gaussian and doughnut-shaped pulses: From localized laser energy absorption to absorption delocalization

Martin Zukerstein[a], Vladimir P. Zhukov[a,b,c], Yuri P. Meshcheryakov[d], Nadezhda M. Bulgakova[a*]
[a]HiLASE Centre, Institute of Physics ASCR, Za Radnici 828, 25241 Dolni Brezany, Czech Republic; [b]Federal Research Center for Information and Computational Technologies, 6 Lavrentyev ave., 630090 Novosibirsk, Russia; [c] Novosibirsk State Technical University, 20 Karl Marx Ave., 630073, Novosibirsk, Russia; [d]Design and Technology Division, Lavrentyev Institute of Hydrodynamics SB RAS, 15 Lavrentyev ave., 630090 Novosibirsk, Russia

**ABSTRACT**

Volumetric modification of glass materials by ultrashort laser pulses is a powerful technique enabling direct writing of three-dimensional structures for fabrication of optical, photonic, and microfluidic devices. The level of modification is determined by the locally absorbed energy density, which depends on numerous factors. In this work, the effect of the spatial pulse shape on the ultrashort laser excitation of fused silica was investigated experimentally and theoretically for the volumetric modification regimes. We focused on two shapes of laser pulses, Gaussian and doughnut-shaped (DS) ones. It was found that, at relatively low pulse energies, in the range of ~1–5 microjoules, the DS pulses are more efficient in volumetric structural changes than Gaussian pulses. It is explained by the intensity clamping effect for the Gaussian pulses, which leads to the delocalization of the laser energy absorption. In the DS case, this effect is overcome due to the geometry of the focused beam propagation, accompanied by the electron plasma formation, which scatters light toward the beam axis. The thermoelastoplastic modeling performed for the DS pulses revealed intriguing dynamics of the shock waves generated because of tubular-like energy absorption. It is anticipated that such a double shock wave structure can induce the formation of high-pressure polymorphs of transparent materials that can be used for investigations of nonequilibrium thermodynamics of warm dense matter. The DS laser pulses of low energies of the order of 100 nJ which generate a gentle tubular-like modification can be perspective for a miniature waveguide writing in glass.

## 1. INTRODUCTION

Localized volumetric modification of transparent materials by ultrashort laser pulses has been demonstrated as a versatile technique used nowadays in many applications such as waveguides [1-3], waveplates [4], Bragg gratings [5-6], optical memories [7-8], and microfluidic devices [9]. To perform volumetric laser processing, the Gaussian pulses are conventionally used. However, it was shown that spatiotemporally shaped pulses can be advantageous for some applications [10-12]. As an example, Bessel laser beams are very efficient for high aspect ratio structuring, drilling, and cutting dielectric materials [13].

Recently, we demonstrated numerically and experimentally that, using doughnut-shaped (DS) laser pulses, it is possible at a certain range of pulse energies to achieve much higher absorption within a strongly localized volume inside fused silica glass as compared with Gaussian pulses [14-15]. This effect can be used for optimizing 3D direct writing as well as for achieving extreme thermodynamic states of matter [16]. In this work, we summarize our further findings and show that, even with very small pulse energies in the DS pulses, it is possible to gently modify glass density for inscription structural changes which can be used in optoelectronic and photonic applications.

*bulgakova@fzu.cz; phone +420 314-007-709; https://www.fzu.cz

## 2. EXPERIMENTAL

The details of the experimental arrangement are described in [15]. Shortly, a Ti:sapphire laser system Astrella (Coherent) was used with a pulse duration of $\tau_{FWHM} \approx 40$ fs and wavelength of 800 nm (see the experimental scheme in [15]). The laser beam was focused inside 1-mm thick plates of fused silica 200 μm below the surface of the sample using an objective with 10× magnification (NA = 0.25). Structural changes within the modified areas were examined with an optical microscope (Olympus BX43) by focusing on a plane of the laser-affected zone having the highest modification contrast. The modifications were studied for accumulated pulses (1; 2; 3; 5; 10; 20; 50; 100; 1000) acting on the same sample region. The main aim here was to compare modifications by Gaussian and doughnut-shaped pulses for verifying our modeling predictions.

## 3. MODEL

Numerical simulations of fused silica irradiation for the conditions used in the experimental studies were based on non-linear Maxwell's equations supplemented by the equations describing the excitation of electrons, their trapping, possible re-excitation, and oscillations of the conduction electrons in the field of the laser wave [14,15,17]. The key equations are as follows:

$$\frac{1}{c}\frac{\partial \mathbf{D}}{\partial t} - i\frac{\omega}{c}\mathbf{D} = \frac{4\pi}{c}e\rho\mathbf{v} + \text{rot }\mathbf{B} - \frac{8\pi}{c|E|^2}\hbar\omega(\alpha W_{PI} + \alpha_{STE}W_{PI}^{STE})\mathbf{E}, \tag{1}$$

$$\frac{1}{c}\frac{\partial \mathbf{B}}{\partial t} - i\frac{\omega}{c}\mathbf{B} = -\text{rot }\mathbf{E}, \tag{2}$$

$$\frac{\partial \rho}{\partial t} = W_{PI} + W_{PI}^{STE} + W_\sigma - \frac{\rho}{\tau_{tr}}, \tag{3}$$

$$i\omega\mathbf{v} = (e/m_e)\mathbf{E} + \frac{\mathbf{v}}{\tau_c}. \tag{4}$$

Here $\mathbf{E}$, $\mathbf{B}$, and $\mathbf{D}$ denote the electric and magnetic fields of the laser wave and the electric displacement field respectively; $\omega$ is the laser frequency at $\lambda = 800$ nm; $c$ is the speed of light; $\rho$ and $\mathbf{v}$ are respectively the density and the velocity of the electrons excited to the conduction band; $e$ is the elementary charge; $W_{PI}$, $W_{PI}^{STE}$, and $W_\sigma$ are the multiphoton ionization coefficients for the valence band electrons and excitons and the collisional ionization rate; $\alpha$ and $\alpha_{STE}$ are the orders of multiphoton ionization from the valence band and exciton states (equal to 6 and 4 respectively for fused silica at 800-nm irradiation); $m_e$ is the electron mass; $\tau_c$ and $\tau_{tr}$ are the electron collision time and the trapping time respectively. The shape of the laser pulse entering the sample at its boundary corresponds to that focused by a parabolic mirror [18,19]. The pulse incident on the mirror reads as

$$\mathbf{E}_{inc} = \mathbf{E}_0(r)\exp(-i\omega(t + z/c) - (t + z/c)^2/t_L^2) \tag{5}$$

where $r$ and $z$ are the beam radius and the direction of the beam propagation. Further details on modeling equations and used parameters can be found in [15]. Here we add specification for the distribution of the absorbed laser energy density as this value determines the material modification level. For its mapping within the laser-affected region in the focal zone, the following integration was performed during the simulations:

$$E_{ab} = \int_{t_0}^{t_1} \left(\frac{\mathbf{jE}^* + \mathbf{j}^*\mathbf{E}}{4} + \alpha\hbar\omega W_{PI} + \alpha_{STE}\hbar\omega W_{PI}^{STE}\right)dt. \tag{6}$$

This integral was calculated for each cell of the numerical grid from the simulation start ($t_0$) till the time moment ($t_1$) when the laser beam left the focal zone and further light absorption became negligible. The numerical scheme and its realization are described in [20].

To gain an insight into the evolution of the laser-excited fused silica under the action stress waves generated due to sudden volumetric heating with the DS pulses, we applied thermoelastoplastic modeling whose details are found in Ref. [21]. The model describes the dynamics of the stress waves and relocation of material with the creation of compacted and rarefied zones. It can also predict the final distribution of the material density for the cases when the material is not melting or melting only very locally (note that the model is based on solid-state hydrodynamics). The model was successfully applied to study the material evolution for the cases of Gaussian [22] and DS [15] ultrashort laser pulses. Here we recall a possible behavior of fused silica at the moderate-energy DS laser pulses and apply the model for low pulse energies when the final modification can be very gentle but can occur to be useful for 3D laser writing of photonic structures.

## 4. RESULTS AND DISCUSSION

Figure 1 presents the optical microscope images of the samples modified by Gaussian (left) and DS (right) laser pulses at different pulse energies (numbers in the columns given in μJ) and at different numbers of pulses coupling to the same region inside the sample (numbers in the bottom rows). It is seen that the DS laser pulses provide a stronger modification compared to the Gaussian ones. Interestingly, that modification with the DS pulses is observable already for single laser pulses with energies ≥ 1.5 μJ. For the Gaussian pulses, some modification becomes visible for 5 pulses or more. As a whole, modification by the DS pulses looks to be stronger compared to the Gaussian pulses that supports our theoretical predictions [14]. Below we shortly repeat our predictions, consider the case of a very low pulse energy, only 100 nJ, and discuss a possible use of such regimes for volumetric material structuring.

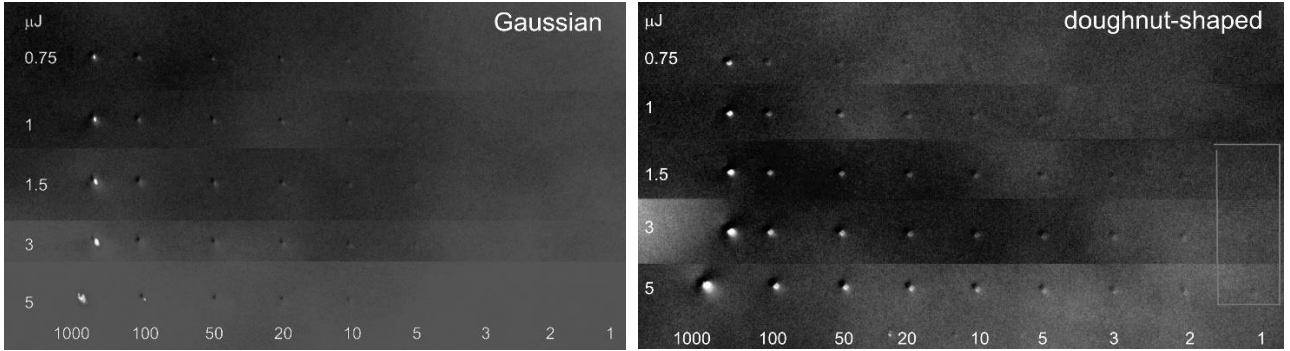

Figure 1. Optical microscope images of the fused silica samples modified by the Gaussian (left) and the doughnut-shaped (right) laser pulses. Structural changes within the modified areas were visualized by focusing on a plane of the laser-affected zone with the highest modification contrast. On the left in the images, the pulse energy is marked in μJ. At the bottom, the number of pulses irradiating the same sample region is indicated.

In Figure 2(a), the map of the absorbed laser energy density is presented for a single DS pulse at the energy of 2 μJ focused to the depth of 120 μm inside the fused silica sample. It demonstrates the extreme effectiveness of the DS pulses as compared to Gaussian pulses. For this particular case, the maximum absorbed energy density is ~14 times higher for the DS pulse than for the Gaussian one [14]. The absorption region has a form of a hollow "nanocylinder" with a cold interior and exterior. The distribution of the absorbed laser energy, being converted to the lattice temperature distribution based on thermodynamic relations [22], can be used as the initial condition for thermoelastoplastic modeling. As can be expected, the hot hollow "nanocylinder" serves as a source of the two shock waves, one outgoing to the periphery and another one converging radially to the beam axis (Figure 2(b)). The converging shock wave, upon cumulation to the axis, creates a high-pressure region where the matter can potentially be brought to extreme states resulting in the formation of high-pressure phases of materials [16].

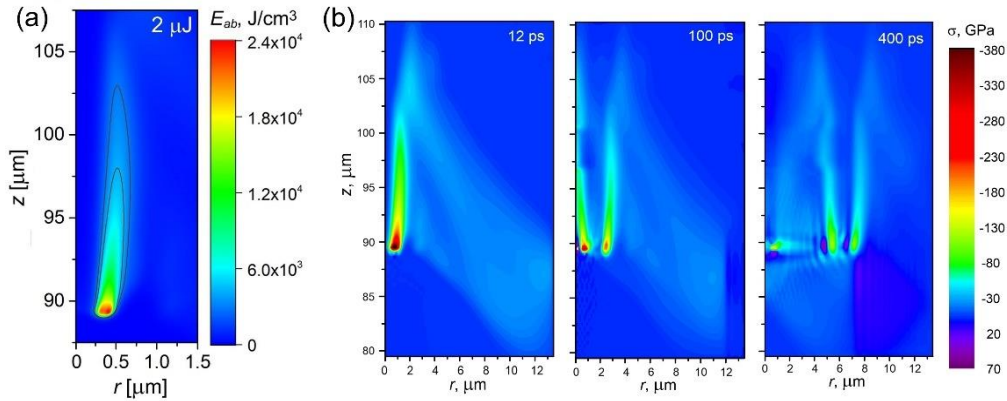

Figure 2. (a) Spatial map of the absorbed laser energy inside fused silica after the action of a 2-μJ DS laser pulse with a duration of 45 fs [14]. The pulse was focused to the sample depth of 120 μm with the lens of NA = 0.25. The laser beam propagates from the bottom. (b) The dynamics of shock waves [15] generated for the absorption conditions of (a).

The stress distribution presented in Figure 2(b) for 12 ps represents a stage when the material has been heated but is not moving yet. According to simulations, the stress achieved at this stage can be of the order of 0.4 TPa. If such a stress level is realized, one can potentially expect the formation of high-pressure material polymorphs transiently [23] or permanently [24]. As seen in the snapshots for 100 and 400 ps, the double shock structure is generated as expected. The shock wave emitted to the "nanocylinder" center has been reflected from it so that the pair of shock waves are propagating to the periphery. At 100 ps, the maximum pressure in the waves is still exceeding 200 GPa. With time, the stress level gradually decreases due to the propagation geometry. However, the pressure exceeds 100-GPa level at 400 ps and drops below 100 GPa only by the time moment of ~650 ps. We note that the shock wave velocity is ~16.5 km/s which agrees well with the experimental data [25]. This indicates that the thermoelastoplastic model applied here is reliable and can be used for investigations of laser-induced shock formation and its propagation. It should be mentioned that the questions about structural transformations under the described extreme conditions are still open. The model cannot address the issue where such transformations are transient or can permanently be imprinted in the sample which calls for further studies.

We also underline that, due to several assumptions made in the model (ionization mechanisms and their rates, electron collision frequency, etc.), the efficiency of the DS beam coupling with the material can differ from that reported here. However, the reported tendency should be general. Also, in such an extreme case, some regions of material should experience melting, vaporization, and even atomization (in Figure 2(a), the solid lines show the energy levels corresponding to annealing (external) and melting (internal) points of fused silica). This also imposes some restrictions on the use of a solid-state model for such localized regions. Below we report the case where the thermoelastoplastic model is applicable and can predict the final density map after laser excitation.

When a low-energy Gaussian laser beam, of the order of 100 nJ, is focused on the depth of fused silica samples, there are no visible signs of the presence of a modified region even at such a high numerical aperture as 0.45 [26]. Here we show that, using DS laser pulses, it is possible to gently induce relocation of the material which, at multi-pulse irradiation, can be suitable for imprinting a new type of modification geometries. Figure 3 presents the simulation results for a 100-nJ DS laser beam action on fused silica under all other conditions as in Figure 2. The left image shows the spatial distribution of the absorbed laser energy density. When converted to the temperature map, the maximum temperature achieved (~2400 K) is still below the boiling point under near-equilibrium conditions [27]. A very tiny region with the dimensions of ~0.14×1.5 μm$^2$ can be temporally converted to a molten but still viscous state. The stress experienced by the material is very moderate with the maximum of the order of -130 MPa. It is achieved at around 100 ps after the laser beam action in the center of the laser-affected region where the stress wave collapses, reflects, and then propagates radially to the sample periphery (Figure 3, right).

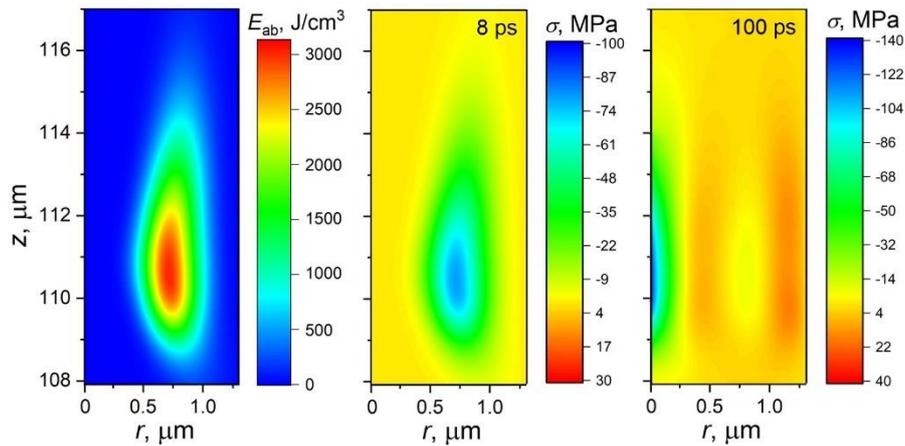

Figure 3. Spatial distribution of the absorbed laser energy inside fused silica after the action of a 100-nJ DS laser pulse (left) with a duration of 45 fs. The focusing conditions are the same as in Figure 2. The middle and right images demonstrate the stress distributions at the initial state in the lase-affected zone and for the moment when the stress wave collapses in the center of this zone. The laser beam propagates from the bottom.

The material is temporally densified in the shock-wave collapse region (Figure 4, left). However, due to relatively moderate stress, material is released back to the initial state (elastic behavior). Only the laser-affected zone, where the material experienced softening and melting, preserves the final imprinting of a slightly lower density in the form of an annular region (Figure 5, right). We anticipate that such irradiation conditions in the multi-pulsed regime can result in the formation of a tubular structure with a core where the refractive index is close to that of the virgin glass surrounded by a cylindrical wall with a lower refractive index. By slow longitudinal beam scanning [11], a miniature waveguide with a diameter of 1–1.5 µm can plausibly be imprinted in glass which calls for further studies.

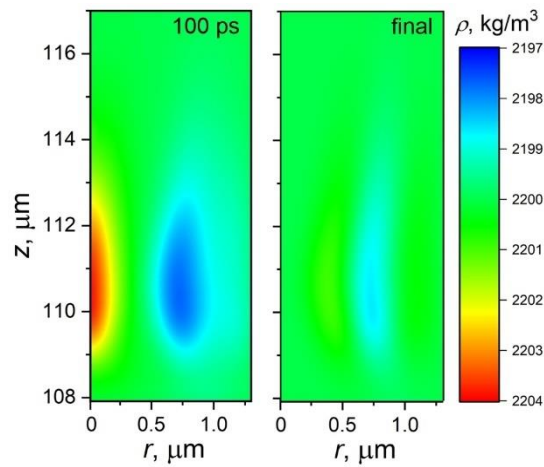

Figure 4. The density distributions inside fused silica after the action of a 100-nJ DS laser pulse at the time moment of 100 ps after the laser beam action (left) and when the final structural change is imprinted in the sample. The irradiation conditions are as in Figure 3.

## 5. CONCLUSIONS

We have investigated experimentally and theoretically the effects of the action of ultrashort laser pulses with the doughnut shape on fused silica in the regimes of volumetric modifications. Experimentally it has been demonstrated that, at moderate pulse energies, in the range of ~1–5 µJ, the DS laser pulses provide stronger structural changes as compared to Gaussian pulses, as was predicted theoretically [14]. To understand the origin of this difference, thermoelastoplastic modeling was performed for the DS pulses. As a result of tubular-like energy absorption, the double shock-wave structure has been formed which leads to the cumulation of material toward the axis of the laser beam propagation. The level of stress can plausibly achieve several hundred GPa. We anticipate that such stresses if achieved can result in the formation of high-pressure polymorphs, transiently or even permanently. Such regimes can also be suitable for the fabrication of microfluidic devices. The DS laser pulses of low energies of the order of 100 nJ which generate a gentle tubular-like modification can be perspective for a miniature waveguide writing in glass.

## ACKNOWLEDGEMENTS

This paper is our tribute to the memory of Dr. Alexander M. Rubenchik, our dear friend, teacher, and outstanding scientist, who inspired this work and provided very helpful comments. This work was supported by EU and MEYS, Project SenDiSo: CZ.02.01.01/00/22_008/0004596.